\RequirePackage{amsthm}

\documentclass[sn-mathphys-ay]{sn-jnl}

\usepackage{graphicx}%
\usepackage{multirow}%
\usepackage{amsmath,amssymb,amsfonts}%
\usepackage{mathrsfs}%
\usepackage[title]{appendix}%
\usepackage[table]{xcolor}%
\usepackage{textcomp}%
\usepackage{manyfoot}%
\usepackage{booktabs}%
\usepackage{algorithm}%
\usepackage{algorithmicx}%
\usepackage{algpseudocode}%
\usepackage{listings}%
\usepackage{makecell}
\usepackage{subfig}
\usepackage{lmodern}
\usepackage{anyfontsize}

\theoremstyle{thmstyleone}%

\theoremstyle{thmstyletwo}%

\theoremstyle{thmstylethree}%

\newcommand{\logdet}{\mathtt{logdet}}
\newcommand{\ySy}{\mathtt{ySy}}
\newcommand{\XSX}{\mathtt{XSX}}
\newcommand{\ySX}{\mathtt{ySX}}
\newcommand{\dlogdet}{\mathtt{dlogdet}}
\newcommand{\dySy}{\mathtt{dySy}}
\newcommand{\dXSX}{\mathtt{dXSX}}
\newcommand{\dySX}{\mathtt{dySX}}
\newcommand{\ainfo}{\mathtt{ainfo}}

\newcommand{\nparms}{\text{nparms}}
\newcommand{\dimv}{d}

\newcommand{\locs}{\mathtt{locs}}
\newcommand{\NNarray}{\mathtt{NNarray}}
\newcommand{\covparms}{\mathtt{covparms}}

\newcommand{\covmat}{\mathtt{covmat}}
\newcommand{\dcovmat}{\mathtt{dcovmat}}
\newcommand{\locsub}{\mathtt{locsub}}
\newcommand{\Xsub}{\mathtt{Xsub}}
\newcommand{\ysub}{\mathtt{ysub}}

\newcommand{\Liy}{\mathtt{Liy}}
\newcommand{\LiX}{\mathtt{LiX}}

\newenvironment{myfont}{\fontfamily{qcr}\selectfont}{\par}
\DeclareTextFontCommand{\textmyfont}{\myfont}

\newcommand{\cc}{\cellcolor[gray]{1}}

\begin{document}

\title[Article Title]{Implementation and Analysis of GPU Algorithms for Vecchia Approximation}

\author*[1]{\fnm{Zachary} \sur{James}}\email{zj37@cornell.edu}

\author[1]{\fnm{Joseph} \sur{Guinness}}\email{guinness@cornell.edu}

\affil[1]{\orgdiv{Department of Statistics and Data Science}, \orgname{Cornell University}, \orgaddress{\city{Ithaca}, \postcode{14853}, \state{NY}, \country{USA}}}

\abstract{Gaussian Processes have become an indispensable part of the spatial statistician's toolbox but are unsuitable for analyzing large dataset because of the significant time and memory needed to fit the associated model exactly. Vecchia Approximation is widely used to reduce the computational complexity and can be calculated with embarrassingly parallel algorithms. While multi-core software has been developed for Vecchia Approximation, such as the \texttt{GpGp} R package, software designed to run on graphics processing units (GPU) is lacking, despite the tremendous success GPUs have had in statistics and machine learning. We compare three different ways to implement Vecchia Approximation on a GPU: two of which are similar to methods used for other Gaussian Process approximations and one that is new. The impact of memory type on performance is investigated and the final method is optimized accordingly. We show that our new method outperforms the other two and then present it in the \texttt{GpGpU} R package. We compare \texttt{GpGpU} to existing multi-core and GPU-accelerated software by fitting Gaussian Process models on various datasets, including a large spatial-temporal dataset of $n>10^6$ points collected from an earth-observing satellite. Our results show that \texttt{GpGpU} achieves faster runtimes and better predictive accuracy.}


\maketitle

\section{Introduction}\label{sec:intro}

Gaussian Processes (GPs) are models that can capture dependencies between observations. These dependencies can be accounted for in regression and then used for interpolation. Furthermore, GPs are highly flexible, provide uncertainty measures, and incorporate domain knowledge through our choice of the mean and covariance functions. As a result, GPs are widely used in spatial statistics, where accounting for the dependence in spatial and spatial-temporal data is crucial, and performing interpolation is a common task.

However, fitting a GP model becomes intractable as the number of data points $n$ increases. Calculating the likelihood involves storing and factoring the covariance matrix, which requires $O(n^3)$ time and $O(n^2)$ memory and cannot be easily adapted to multi-core architectures. As a result, fitting a GP via direct calculation becomes infeasible when $n$ is large, such as for $n > 10^5$. This is problematic because advances in remote sensing technology have made larger dataset ever more common. For example, modern satellite sensors now have resolutions that allow for millions of observations on a single orbit.

Approximation methods address this computational problem by simplifying the linear algebra operations in the likelihood calculation. Inducing point methods assume that a subset of the data can be used to capture the dependency structure, allowing the covariance to be expressed in terms of low-rank matrices \citep{Rasmussen2005, Banerjee2008}. Covariance tapering modifies the covariance function with a compact positive semi-definite function to induce sparsity in the covariance matrix \citep{sparsecov06, sparsecov08}. Several methods can be adapted to produce sparse precision matrices, including Bayesian inference \citep{Tan2018} and the multitude of methods that use the stochastic partial differential equation approximation \citep{lindgren2021}. All these methods eliminate the need to store a dense $O(n^2)$ covariance matrix and, in turn, reduce the time and memory burden.

The development of approximation methods has happened concurrently with advances in computing hardware, specifically more powerful central processing units (CPUs) and graphics processing units (GPUs). The release of open-source Message Passing Interface (MPI) libraries has made it significantly easier to run programs on large-scale multi-core CPU and GPU computing systems, while newer frameworks, such as OpenMP, make running algorithms in parallel as simple as adding a handful of directives to the source code. GPUs, which were initially developed for graphics rendering, have been used to great success in the machine learning and scientific computing communities with the help of CUDA, OpenCL, and other frameworks. 

As a result, practitioners have used GP approximations in conjunction with advanced hardware to analyze large datasets. Early attempts simply fit exact methods or existing approximations on these systems \citep{franey2012short, Abdulah2018}, often focusing on computing one aspect of the likelihood, such as the Cholesky decomposition, in parallel \citep{bigGP}. However, many of the required linear algebra subroutines, including the Cholesky decomposition, cannot be easily computed on parallel systems, resulting in minimal speed-up.

One solution is to use approximation methods that can be calculated in parallel. Local approximate GP \citep{laGP} fits separate models for each prediction point by using only nearby observations for each model. Since the local models are independent of one another, the method can be fit with embarrassingly parallel algorithms. There has also been success in adapting Bayesian GP methods to parallel computing by expressing the lower bound in variational inference as a sum over the data points and then calculating it in parallel \citep{GPy, GPflow}. Other methods try to approximate specific terms in the exact likelihood of the GP using linear algebra subroutines designed for parallel systems \citep{GPytorch}. An overview of software for fitting GP models is provided in Section \ref{sec:existingsoftware}.

Vecchia Approximation \citep{vecchia88} is a popular approximation that assumes observations are independent of one another when conditioned on nearby observations. It is highly accurate in terms of Kullback–Leibler divergence \citep{guinness2018perm} and defines a valid marginal distribution \citep{Datta2016}. Computationally, it replaces the storing and factoring of the $n\times n$ covariance matrix of the marginal distribution with $n$ matrices of size $O(m\times m)$ for each conditional distribution, where $m$ is a hyperparameter chosen to be much smaller than $n$. Each of these conditional distributions can then be computed in parallel, which existing software exploits by using multi-core CPU computing. \citep{pan2024} made an important contribution by demonstrating how to calculate the likelihood function on a GPU with batched methods. However, it is unclear how to best optimize the likelihood for parameter estimates on a GPU.

We close this gap by creating a software package that fits a GP with Vecchia Approximation on an NVIDIA GPU. We use the decompositions in \citep{guinness2019fisher} to quickly calculate the likelihood, gradient, and Fisher information of a GP model, and then use Fisher scoring to find parameter estimates. We investigate three different methods for implementing the \citep{guinness2019fisher} algorithm on a GPU, two of which are similar to other methods used for GP approximation software, and one that is novel. We find that our novel method is superior and implement it in a software package for R.

The final method relies on several optimizations to significantly reduce run times. Since each parallel process of Vecchia Approximation has a low memory complexity, we can store data in registers, which have the lowest read and write latency. This is validated by comparing our method to the other two methods, which rely on global and shared memory. We also nearly eliminate the need to synchronize the threads, the most basic computing units on the GPU, allowing us to calculate the likelihood of a GP on more than a million points fully in parallel.

The remainder of this paper is organized as follows. In Section \ref{sec:background} we describe Vecchia Approximation, introduce the reader to the architecture of a GPU, and provide an overview of existing software for fitting GP models. In Section \ref{sec:methodology} we introduce our software and describe the three different implementations. In Section \ref{sec:numerical} we present a numerical study comparing the performance of our methods to the \texttt{GpGp} software in R, which uses multi-core parallelization, and \texttt{GPytorch} in Python, which uses a GPU. Finally, in Section \ref{sec:discussion} we conclude with a discussion and possible directions for future work. The methods described in this paper are implemented in the R package \texttt{GpGpU}, which is available at \url{https://github.com/zjames12/GpGpU}. 

\section{Background}\label{sec:background}

\subsection{Vecchia Approximation}\label{sec:vecchia}

A GP is a stochastic process whose finite dimensional realizations follow multivariate normal distributions. Its covariance structure allows for the dependencies between observations to be modeled. Given response vector $Y=(Y_1,\ldots,Y_n)^T$ and covariates $X=(X_1,\ldots,X_n)^T$ observed at fixed locations $Z=(Z_1,\ldots,Z_n)^T$, we assume a GP process $Y\sim N(\mu(X,Z,\beta),\Sigma(X, Z, \theta))$ with mean parameters $\beta$ and covariance parameters $\theta$. Without loss of generality, let $Z_i\in\mathbb{R}^d$. We will also assume that the mean process is linear $\mu(X,Z,\beta)=X\beta$ and the covariance is independent of the attributes $\Sigma(X, Z, \theta)=\Sigma(Z, \theta)$. Since the locations are fixed, denote $\Sigma(Z, \theta)$ as $\Sigma(\theta)$.

GP models have been effectively used to analyze spatial data. They are highly expressive and can model any smooth spatial process arbitrarily well \citep{Micchelli2006}. Regression parameters can be profiled given covariance parameter estimates. They can be used to interpolate the response at an unobserved location, a process known as kriging. Obtaining prediction intervals for the interpolations is straightforward.

The practical application of GP models is hindered by the cost of computing the log-likelihood and score functions:
\begin{equation}
    \mathcal{L}(\theta,\beta;Y,X) = -n\log 2\pi-\log\det\Sigma(\theta)-(Y-X\beta)^T\Sigma(\theta)^{-1}(Y-X\beta)
\end{equation}
\begin{equation}
    \frac{\partial\mathcal{L}}{\partial\theta} = -\text{Tr}\left(\Sigma(\theta)^{-1}\frac{\partial\Sigma(\theta)}{\partial\theta}\right) + (Y-X\beta)^T\Sigma(\theta)^{-1}\frac{\partial\Sigma(\theta)}{\partial\theta}\Sigma(\theta)^{-1}(Y-X\beta),
\end{equation}
which scale poorly with the number of observations $n$. In particular, it involves the log-determinant of the covariance matrix $\log\det\Sigma(\theta)$, inverse of the covariance matrix $\Sigma(\theta)^{-1}$, and trace of the covariance multiplied by its derivative $\text{Tr}\left(\Sigma(\theta)^{-1}\frac{\partial\Sigma(\theta)}{\partial\theta}\right)$. Computing these quantities requires $O(n^3)$ time and $O(n^2)$ space. This quickly becomes infeasible on most computers as $n$ increases.

Vecchia Approximation addresses these computational issues by decomposing the marginal distribution into approximate conditional distributions \citep{vecchia88}. First, an ordering is established on the realized observations $y=\{y_1,\ldots,y_n\}$. For each $y_i$ we find the set $s_i\subseteq \{y_{1},\ldots y_{i-1}$\} of the $\min(i-1,m)$ points that precede $y_i$ in the ordering and have the locations that are closest to $y_i$'s location, where $m$ is a hyperparmeter selected to be much smaller than $n$. We then assume that each observation $y_i$ conditioned on $s_i$, referred to as the nearest neighbors, is independent of all other observations. This expresses the density
\begin{align}
    P(Y=y) &= P(y_1)\prod_{i=2}^n P(y_i\mid y_1,\ldots,y_{i-1})    \approx  P(y_1)\prod_{i=2}^n P(y_i\mid s_i)
\end{align}
as the product of approximate conditional densities.

The approximation significantly reduces the computational burden by lowering the time and memory complexity to $O(nm^3)$ and $O(nm^2)$ respectively. The conditional likelihoods can be computed in parallel, involve matrices of size at most $(m+1)\times(m+1)$, and only depend on a small subset of the data. \citep{guinness2019fisher} showed how the corresponding score function and Fisher information can also be calculated in parallel.

Vecchia Approximation has several desirable properties. The ordering ensures that the full likelihood is returned when $m=n$, and for $m<n$ the conditional distributions imply a valid marginal distribution that has a covariance matrix with a sparse inverse Cholesky factor. The approximation is optimal in terms of the Kullback–Leibler divergence between the true and implied distributions \citep{katzfuss2021}. Calculating the maximum likelihood estimates is equivalent to solving a set of unbiased estimating equations \citep{stein2004}.

Numerous extensions to Vecchia Approximation have been proposed. The ordering of the points significantly impacts the parameter estimates, and heuristic-based ordering algorithms have been developed \citep{guinness2018perm}. While Vecchia considered conditioning on individual observations, we can also condition on joint distributions. This grouping reduces the number of likelihood calculations while improving the accuracy of the approximation \citep{guinness2018perm}. There are also considerable variations on how the conditioning sets are chosen. We may wish to include far-away observations to capture how the spatial process decays \citep{stein2004}, or the conditioning sets may be learned \citep{Datta2016}.

\subsection{GPU Computing}
\label{sec:gpucomp}

A GPU is a type of computer processor originally designed for rendering computer graphics, which involves manipulating images stored as arrays. GPUs can perform a large number of tasks in parallel by using  specialized hardware design to generate separate computational processes for each task.

The dynamic parallelism of GPUs has made them ideal for highly parallel tasks unrelated to graphics processing, especially in statistics and machine learning. Many linear algebra subroutines experience significant speedup when ported on a GPU, such as the matrix-matrix multiplies needed to train neural networks. Other applications include principal component analysis \citep{Andercut2009} and Markov chain Monte Carlo sampling \citep{Beam2016}.

Much of the progress in GPU computing has been facilitated by the CUDA Toolkit from NVIDIA, which provides resources for writing C/C++ code for NVIDIA GPUs. In the CUDA programming paradigm, users write \textit{kernels} (unrelated to kernels in the GP literature) that run on the GPU. Each kernel initializes a \textit{grid},  a multidimensional array of \textit{blocks}, each of which is a multidimensional array of \textit{threads}, a single computational process. The grid and blocks can be created as three-dimensional arrays so that a kernel with grid dimension $10\times 10\times10$ and block dimension $10\times 10\times10$ would have $10^6$ threads. We can also choose to either synchronize all the threads we create, or just those in a specific block, creating two levels of parallelism.

How we use this parallelism impacts the performance of the kernel. Excessive synchronization can have a negative impact, especially if it results in the computational load being unequally distributed among the threads. It can also impact how we use GPU memory, which is separate from CPU memory and has different properties depending on the type. Global memory can be accessed by any process on the GPU, but has high read and write latency. Shared memory can be accessed by all the threads in the same block, but is limited in size. Registers have the lowest access time but are also small in size. Developers exploit different types of parallelism and memory when writing kernels to reduce computation time.

NVIDIA has also developed its own kernels and has presented them in the cuBLAS and cuSOLVER libraries. Similar to the well-known BLAS and LAPACK packages, these libraries provide common linear algebra subroutines that are designed to run on a GPU. Developers have found success using these functions as drop-in replacements for sequential linear algebra subroutines.

\subsection{Existing Software}
\label{sec:existingsoftware}

There are many software libraries for fitting GP models, a large number of which support parallel computing. We will summarize GPU software for GP and then examine the few existing software programs for computing Vecchia Approximation on parallel systems.

Early work fitting GPs on GPUs focused on speeding up linear algebra operations such as calculating covariance matrices and finding LU decompositions \citep{franey2012short}. Particular attention was given to the Cholesky decomposition as it is often the bottleneck in likelihood evaluation and cannot be easily calculated in parallel. The  \texttt{bigGP} package \citep{bigGP} used a block Cholesky algorithm to improve runtimes. The \texttt{GPytorch} package \citep{GPytorch}, on the other hand, eliminates the Cholesky decomposition altogether and instead uses a custom Krylov subspace algorithm to calculate the most expensive terms in the likelihood and gradient. Performing linear algebra operations in parallel reduces each likelihood evaluation to a sequence of parallel processes, often requiring synchronization after each one.

An alternative approach is to use approximation methods to rewrite the likelihood and gradient of the GP as distinct processes that can be computed in parallel. \texttt{GPy} \citep{GPy} expresses the variational lower bound of a GP likelihood as a sum over the data points. However, only two terms in the bound are calculated using the GPU, and it requires thread synchronization within blocks. The \texttt{GPflow} package \citep{GPflow} addresses this by using the GPU to sum all the terms in the variational lower bound. Similarly, the \texttt{laGP} package \citep{laGP} provides an algorithm for the local approximate GP method that can be computed in parallel across the data points. Each data point is assigned a block and an expensive subroutine is then computed in parallel across threads. All of these methods express the underlying process as completely parallel, enabling more efficient use of parallel hardware.

There is limited software for fitting a GP with Vecchia Approximation. The \texttt{GPVecchia} package \citep{GPVecchia} uses OpenMP to calculate the sparsity structure of the inverse Cholesky factor of the covariance matrix. The \texttt{GpGp} package \citep{gpgp} goes further and uses OpenMP to calculate the likelihood, gradients, and Fisher information in parallel. \citep{pan2024} calculates the likelihood of a mean-zero GP with GPU-accelerated batched methods, but does not discuss how to calculate the associated maximum likelihood estimates. Parameter estimation is possible using a gradient-free optimizer, but such optimizers have been found to be inefficient when fitting GP models \citep{guinness2019fisher}. The code also does not support linear predictors or unstructured noise, often refered to as the nugget effect.

\section{GPU Algorithms for Vecchia Approximation \label{sec:methodology}}

We will fit a GP model with Vecchia Approximation on a GPU by using Fisher scoring to calculate maximum likelihood estimates for the covariance parameters $\theta$ and then profiling the mean parameters $\beta$. Fisher scoring iteratively optimizes a likelihood function $\mathcal{L}$ using the score function $\frac{\partial \mathcal{L}}{\partial \theta}$ and Fisher information $\mathcal{I}$.
\begin{equation}
    \theta_{k+1}=\theta_k+\mathcal{I}^{-1}(\theta_k)\frac{\partial \mathcal{L}}{\partial \theta}(\theta_k)
\end{equation}

\citep{guinness2019fisher} shows how to calculate these quantities in parallel, as partially described in Algorithm 1, which takes the response vector, design matrix, locations, covariance parameters, and nearest neighbors as inputs. The nearest neighbors are represented as an $n \times (m+1)$ array of indices, with row $i$ containing the indices of the $\min(i,m+1)$ observations that are closest to observation $i$, including $i$ itself. A separate function takes the output of Algorithm 1 and produces the likelihood, gradient, and Fisher information. The \texttt{GpGp} package implements Algorithm 1 with support for multi-core computing.

We will consider three different GPU implementations for the same algorithm. The first, thread-per-observation, is new, while the other two, block-per-observation and batched, are similar to methods used for fitting other GP approximations. All three methods are implemented in R and use \texttt{Rcpp} \citep{Rcpp} to call C++ functions, which in turn call CUDA functions. All calculations are performed with double precision arithmetic.

\begin{algorithm}
\caption{Sequential Vecchia from \citep{guinness2019fisher}. Calculates several values ($\logdet, \ySy\ldots$) that can be combined to produce the likelihood, gradient, and Fisher information of the GP.}\label{alg:sequential}
\begin{algorithmic}[1]
\State \textbf{Input:} response $y[n]$, design matrix $X[n][p]$, locations $\locs[n][\dimv]$, nearest neighbor indices $\NNarray[n][m+1]$, covariance parameters $\covparms[\nparms]$
\State \textbf{Initialize:} $\logdet,\ySy,\XSX[p][p],\ySX[m+1][\nparms]$ 
\Statex $\dlogdet[\nparms],\dySy[\nparms],\dXSX[p][p][\nparms],\dySX[p][\nparms]$
\Statex $\ainfo[\nparms][\nparms]$
\State \textbf{Initialize:} $\covmat[m+1][m+1], \dcovmat[\nparms][m+1][m+1]$
\Statex $\locsub[m+1][\dimv], \Xsub[m+1][p], \ysub[m+1],\ldots$
\For{$i=1\ldots n$}
\State Substitute nearest neighbors for location $i$ into $\locsub$
\State Substitute corresponding values into $\Xsub$ and $\ysub$.

\State Compute covariance matrix $\covmat$
\State Compute Cholesky of $\covmat$ in-place
\State $\Liy \gets \covmat^{-1}\ysub$
\State $\LiX \gets \covmat^{-1}\Xsub$
\State $\logdet[i]\gets 2\log \covmat[m][m]$
\State $\ySy\gets\ySy + \Liy[m]^2$
\State $\XSX\gets \XSX + \LiX[m,](\LiX[m,])^T$
\State $\ySX\gets\ySX + \Liy[m](\LiX[m,])^T$
\State Compute covariance matrix derivative $\dcovmat$
\State $\ldots$
\Comment{Remaining operations}
\EndFor
\State \Return $\logdet, \ySy, \XSX, \ySX, \dXSX, \dySy, \dySX, \dlogdet, \ainfo$
\end{algorithmic}
\end{algorithm}

\subsection{Thread-per-Observation}
\label{sec:thread}

A straightforward approach is to initialize a kernel with $n$ threads, each of which performs one iteration of the for-loop on line 4 of Algorithm \ref{alg:sequential}. The first $m$ iterations are calculated separately on the CPU to balance the computational load. Since the calculations for each observation are limited to a thread, we can store the covariance matrix, its derivative, and other quantities in registers. Arrays stored in registers cannot be dynamically allocated, so we use C++ template programming to create multiple kernels with different array sizes. The kernel is described in Algorithm \ref{alg:thread}. 

Algorithm \ref{alg:sequential} sums objects over the $n$ data points, which can cause memory errors on parallel systems. Algorithm \ref{alg:thread} ensures thread safety by initializing the objects as arrays in global memory with an extra dimension of length $n$. The results of each parallel process are first written to the array, then transferred to the CPU and summed along the extra dimension to produce the final result. For example, the $\XSX$ quantity, which is initialized on line 2 of Algorithm \ref{alg:sequential} as a $p\times p$ array, is initialized as a $n\times p\times p$ array in Algorithm \ref{alg:thread}. After the kernel executes, the array is transferred to the CPU and summed over the first dimension. While the array could be summed in parallel on the GPU, we find this gives minimal speed-up. This approach results in the intermediary quantities ($\ySy$, $\XSX$, etc.) taking $O(np^2\nparms)$ space and may not be ideal for GP with many attributes or covariance parameters.

The thread-per-observation method has several benefits. Absent memory constraints, the number of observations the algorithm can support is only limited by the maximum number of threads, approximately 67 million for CUDA 12. Single-thread parallelization also reduces memory access times by using registers. Although the size of the registers on each thread is small, the matrices involved in each calculation are, by design, small enough to be stored on them. Finally, the method requires only a single synchronization, to make sure the kernel has finished executing.

Unlike the other two methods we will discuss, Algorithm \ref{alg:thread} does not fully exploit the high degree of parallelism offered by the GPU, as multiple linear algebra operations must be computed sequentially on a single thread. However, the most time consuming computation is the Cholesky decomposition, which is not amenable to parallelization. Furthermore, the other linear algebra operations, such as solving small triangular systems, are fairly simple and can be efficiently performed on a single thread.

\begin{algorithm}
\caption{Kernel for thread-per-observation Vecchia. Calculates several values that can be combined to produce the likelihood, gradient, and Fisher information of the GP.}\label{alg:thread}
\begin{algorithmic}[1]
\State \textbf{Input:} $y[n]$, $X[n][p]$, $\locs[n][\dimv]$  \Comment{Global Memory}
\Statex $\NNarray[n][m+1]$, $\covparms[\nparms]$
\Statex $\logdet[n],\ySy[n],\XSX[n][p][p],\ySX[n][m+1][\nparms]$
\Statex covariance function $cov$, derivative function $dcov\ldots$
\State \textbf{Initialize:} $\covmat[m+1][m+1]$
\Comment{Registers}
\Statex $\dcovmat[\nparms][m+1][m+1]$
\Statex $\locsub[m+1][\dimv], \Xsub[m+1][p], \ysub[m+1]$

\State $i= \text{blockIdx.x} \times \text{blockDim.x} + \text{threadIdx.x}$
\Comment{Thread index}
\If{$i < m$}
\State \Return
\EndIf
\For{$j=1,\ldots,m+1$}
\For{$k=1,\ldots,d$}
\State $\locsub[j][k]=\locs[\NNarray[i][j]][k]$
\EndFor
\For{$k=1,\ldots,p$}
\State $\Xsub[j][k]=X[\NNarray[i][j]][k]$
\EndFor
\State $\ysub[j]=y[\NNarray[i][j]]$
\EndFor

\For{$p=1,\ldots,m+1$ and $q=1,\ldots,p$}
\State $\covmat[p][q] = cov(\covmat, \locsub, \covparms)$
\EndFor
\State Compute Cholesky of $\covmat$ in-place
\State $\Liy \gets \covmat^{-1}\ysub$
\State $\LiX \gets \covmat^{-1}\Xsub$
\State $\logdet[i]\gets 2\log \covmat[m][m]$
\State $\ySy[i]\gets \Liy[m]^2$
\State $\XSX[i,]\gets \LiX[m,](\LiX[m,])^T$
\State $\ySX[i,]\gets \Liy[m](\LiX[m,])^T$
\For{$r=1,\ldots,\nparms,p=1,\ldots,m+1,q=1,\ldots,p$}
\State $\dcovmat[r][p][q] = dcov(\covmat, \locsub, \covparms)$
\EndFor
\State $\ldots$
\Comment{Remaining operations}
\end{algorithmic}
\end{algorithm}

\subsection{Block-per-Observation}
\label{sec:block}
Operations on a GPU can first be divided among blocks and then further divided among threads, providing two layers of parallelism. The threads within a block can use shared memory and block-level synchronization to communicate. This can greatly accelerate tasks with a hierarchical parallel nature, such as the implementation of local approximate nearest neighbors in \texttt{laGP}.

We use this approach by creating a kernel that initializes $n$ blocks, each containing a two-dimensional $(m+1) \times (m+1)$ array of threads. The $n$ iterations of the outer for-loop, lines 4 to 25 in Algorithm \ref{alg:sequential}, are performed in parallel across the blocks, and the operations within each iteration are performed in parallel across the threads. The substitution of nearest neighbors, calculation of covariance matrix, and calculation of covariance derivative are done in parallel, while the remaining operations lack parallel structure and are faster when calculated sequentially. The kernel is described in detail in Algorithm \ref{alg:block}. As in Algorithm \ref{alg:thread} the first $m$ iterations are calculated on the CPU. 

Delegating one block for each observation fully exploits the parallel nature of a GPU and significantly reduces the time needed to calculate the covariance matrix and its derivative. However, it is limited by extensive thread synchronization and an imbalanced computational load. Furthermore, CUDA is only able to generate a limited number of blocks, approximately 65 thousand for CUDA 12. As a result, a block-per-observation implementation cannot be easily scaled to large datasets, and would require either multiple kernels running on separate streams, multiple GPUs, or multiple iterations of the kernel.

\begin{algorithm}
\caption{Kernel for block-per-observation Vecchia. Calculates several values that can be combined to produce the likelihood, gradient, and Fisher information of the GP.}\label{alg:block}
\begin{algorithmic}[1]
\State \textbf{Input:} $y[n]$, $X[n][p]$, $\locs[n][\dimv]$, $\NNarray[n][m+1]$ \Comment{Global Memory}
\Statex $\covparms[\nparms],\logdet[n],\ySy[n],\XSX[n][p][p]$
\Statex covariance function $cov$, derivative function $dcov\ldots$
\State \textbf{Initialize:} $\covmat[m+1][m+1]$
\Comment{Shared Memory}
\Statex $\dcovmat[\nparms][m+1][m+1],\locsub[m+1][\dimv]\ldots$

\State $i = \text{blockIdx.x}$
\Comment{Block index}
\State $j = \text{threadIdx.x}$
\Comment{Thread index first dimension}
\State $k = \text{threadIdx.y}$
\Comment{Thread index second dimension}

\If{$i < m$}
\State \Return
\EndIf
\If{$j < m \text{ and } k < \dimv$}
    \State $\locsub[j][k]=\locs[\NNarray[i][j]][k]$
\EndIf
\If{$j < m \text{ and } k < p$}
    \State $\Xsub[j][k]=X[\NNarray[i][j]][k]$
\EndIf
\If{$j < m \text{ and } k = 0$}
\State $\ysub[j]=y[\NNarray[i][j]]$
\EndIf
\State \textmyfont{\_\_syncthreads()}
\If{$j < m \text{ and } k < j$}
\State $\covmat[p][q] = cov(\covmat, \locsub, \covparms)$
\EndIf
\If{$j < m \text{ and } k < j$}
\For{$r=1,\ldots,\nparms$}
\State $\dcovmat[r][j][k] = dcov(\dcovmat, \locsub, \covparms)$
\EndFor
\EndIf
\State \textmyfont{\_\_syncthreads()}
\If{$j = 0 \text{ and } k = 0$}
    \State \ldots \Comment{Remaining operations on single thread}
\EndIf
\end{algorithmic}
\end{algorithm}

\subsection{Batched Methods}
\label{sec:batch}
Batched methods solve many small linear algebra problems together. Many linear algebra libraries provide batched algorithms for multi-core and GPU systems.  \citep{pan2024} used the KBLAS-GPU library to compute the likelihood of a zero-mean GP with Vecchia Approximation.

We use batched methods by creating kernels to perform some of the operations and then rely on methods from NVIDIA libraries for the rest. Our kernels, shown in Algorithm \ref{alg:batch}, substitute the nearest neighbors, calculate the covariance, and calculate the covariance derivative for all $n$ iterations in parallel. The batched methods from the cuSOLVER and cuBLAS libraries are then used for the Cholesky decomposition and triangular solves.

Batched algorithms allow for efficient memory read patterns. However, they require intermediary quantities to be stored in global memory, which has high read latency. We also need to synchronize the system after nearly every operation, which significantly slows down computation.

\begin{algorithm}
\caption{Batched Vecchia. Calculates several values that can be combined to produce the likelihood, gradient, and Fisher information of the GP.}\label{alg:batch}
\begin{algorithmic}[1]
\State \textbf{Input:} response $y[n]$, design matrix $X[n][p]$  \Comment{Global Memory}
\Statex locations $\locs[n][\dimv]$, nearest neighbor indices $\NNarray[n][m+1]$
\Statex covariance parameters $\covparms[\nparms]$
\State \textbf{Initialize:} $\logdet,\ySy,\XSX[p][p]\ldots$
\Comment{Global memory}
\State \textbf{Initialize:} $\covmat[n][m+1][m+1]$
\Comment{Global Memory}
\Statex $\dcovmat[n][m+1][m+1][\nparms]$
\Statex $\locsub[n][m+1][d], \Xsub[n][m+1][p], \ysub[n][m+1]$
\State substituteNearestNeighborsBatched($\NNarray,\locs,\locsub$)
\State substituteXBatched($\NNarray,X,\Xsub$)
\State substituteyBatched($\NNarray,y,\ysub$)
\State \textmyfont{cudaDeviceSynchronize()}
\State covarianceBatched($\covmat, \locsub, \covparms$)
\State covarianceDerivativeBatched($\dcovmat,\locsub,\covparms$)
\State \textmyfont{cudaDeviceSynchronize()}
\State cusolverDnDpotrfBatched($\covmat\ldots$)
\State \textmyfont{cudaDeviceSynchronize()}
\State cublasDtrsmBatched($\covmat, \ldots$)
\State \textmyfont{cudaDeviceSynchronize()}
\State cublasDtrsmBatched($\covmat, \Xsub$)
\State \textmyfont{cudaDeviceSynchronize()}
\State cublasDtrsmBatched($\covmat, \ysub$)
\State \textmyfont{cudaDeviceSynchronize()}
\For{$i=m\ldots n$}
\State $\logdet\gets \logdet + 2\log \covmat[i][m][m]$
\State $\ySy\gets \ySy + \ysub[i][m]^2$
\State $\XSX\gets \XSX + \Xsub[i][m,](\Xsub[i][m,])^T$
\State $\ySX\gets \ySX + \ysub[i][m](\Xsub[i][m,])^T$
\EndFor
\State $\ldots$\Comment{Remaining operations}
\end{algorithmic}
\end{algorithm}

\clearpage

\section{Numerical Studies}
\label{sec:numerical}

We compare the three GPU implementations discussed Section 3 by fitting a GP model on a real world dataset. The methods return the same parameter estimates and are only compared by wall clock time. The fastest method, thread-per-observation, is presented in the \texttt{GpGpU} package and then compared to the multi-core implementation in \texttt{GpGp} and two different approximation methods in the \texttt{GPytorch} package.

\begin{figure}[h!]
\centering
\includegraphics[scale=0.6]{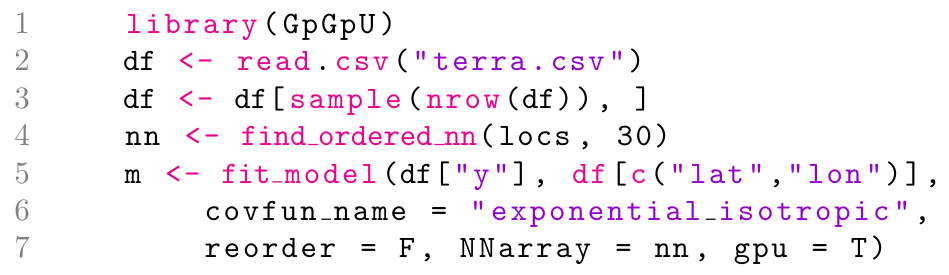}
\caption{Example usage of \texttt{GpGpU} for fitting a GP model on the Terra dataset with $m=30$ neighbors.}\label{fig:code}
\end{figure}

When fitting a model with \texttt{GpGpU} in R, the user must load the data into the R environment, reorder the data, find the nearest neighbors, and fit the model. In the example code in Figure \ref{fig:code} the data is loaded on line 2, points are reordered on line 3, the nearest neighbors are found on line 4, and the model is fit on line 5, which returns the final parameter estimates. \texttt{GpGpU} only applies GPU acceleration to fitting the model, which is usually the most expensive step. We measure the time to fit a model by timing line 5, which includes all the overhead for computing on a GPU, such as allocating memory, transferring data to the GPU, and transferring results back to the CPU.

\subsection{Resources}
\label{sec:resources}
All computations are run on the G2 computing cluster at Cornell University. GPU-accelerated software used a single NVIDIA A40 Tensor Core GPU with compute capability 6.1. The multithreaded CPU software was run on six Intel Ice Lake processors and used the OpenBLAS linear algebra library to achieve faster results.
The R packages are built and run on R 4.3 and the Python packages run on Python 3.8.

\subsection{Data}
\label{sec:data}
We use three real-world datasets, two of which are large geospatial datasets collected by earth-observing satellites, and one that is an experimental dataset previously used to measure the performance of machine learning methods. 

\begin{figure}[h!]
\centering
\includegraphics[scale=0.38]{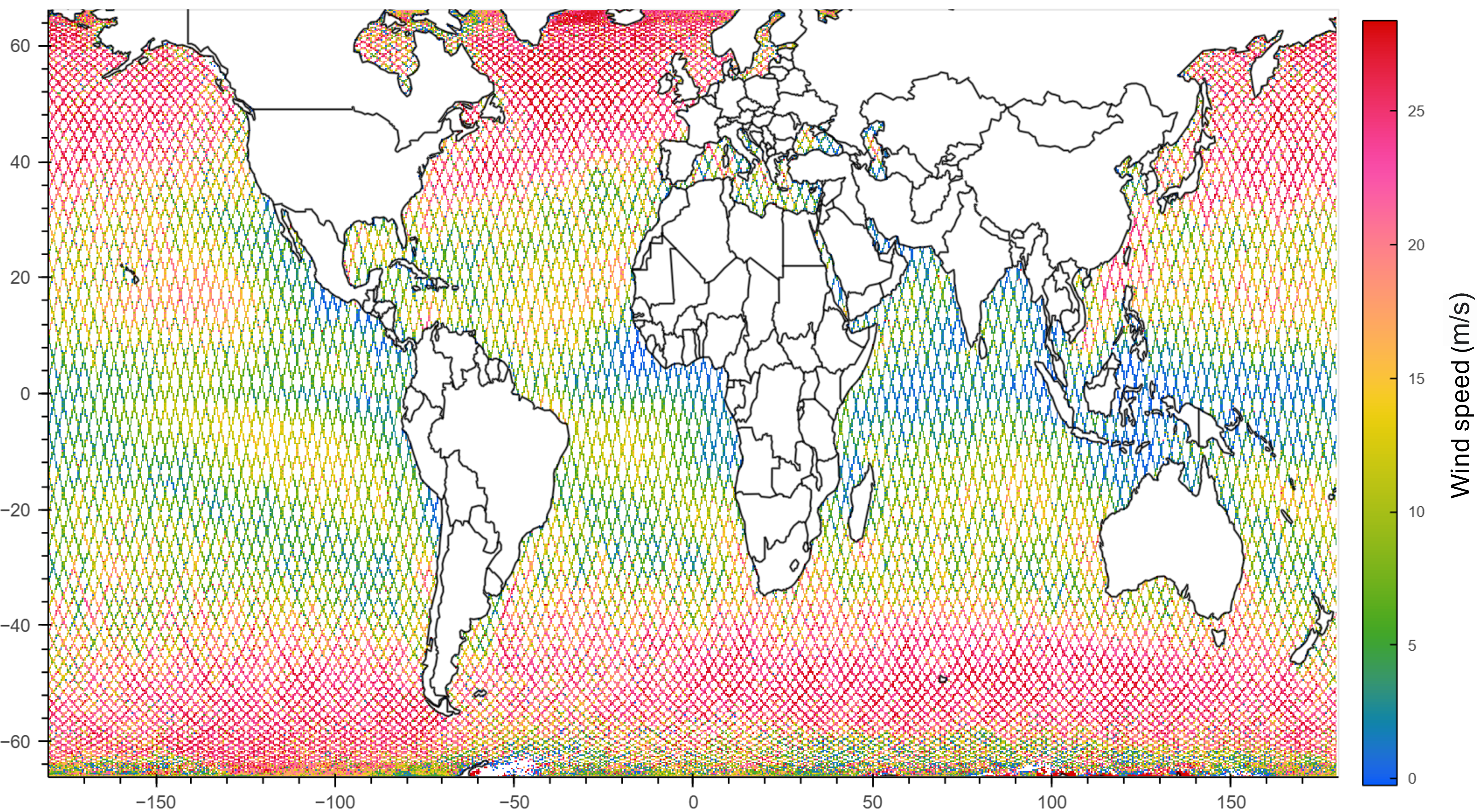}
\caption{Jason-3 ocean surface wind speed readings from \citep{Bekerman2023}.}\label{fig:jason3}
\end{figure}

Jason-3 is an Earth-observing satellite operated by the National Oceanic and Atmospheric Administration (NOAA) and the European Organisation for the Exploitation of Meteorological Satellites (Eumetsat) that uses a radar altimeter for ocean surface readings. \citep{Bekerman2023} used global Jason-3 readings of ocean surface wind speeds to examine biases in satellite sensors. We use a subset of this data from September 18, 2019, to February 29, 2020  and downsample it every eight seconds, resulting in 1,040,815 observations. We also consider a subset of the first 50,000 observations. The data is plotted in Figure \ref{fig:jason3}.

\begin{figure}[h!]
\centering
    \subfloat[Full data]{\includegraphics[scale=0.4]{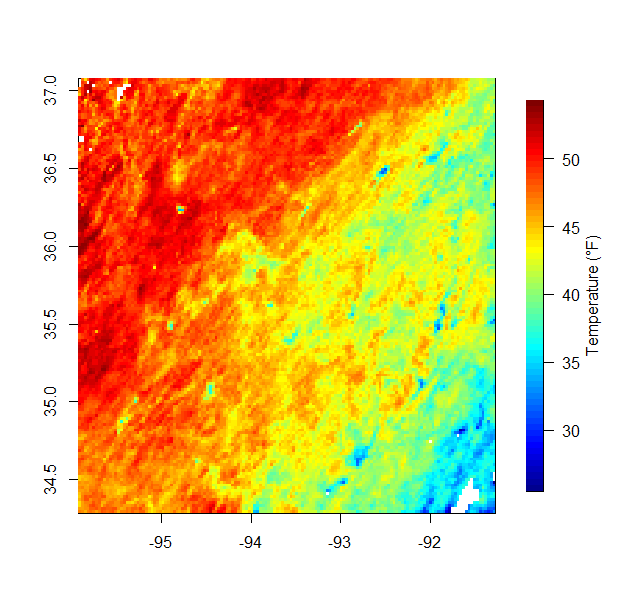}} 
    \subfloat[Training data]{\includegraphics[scale=0.4]{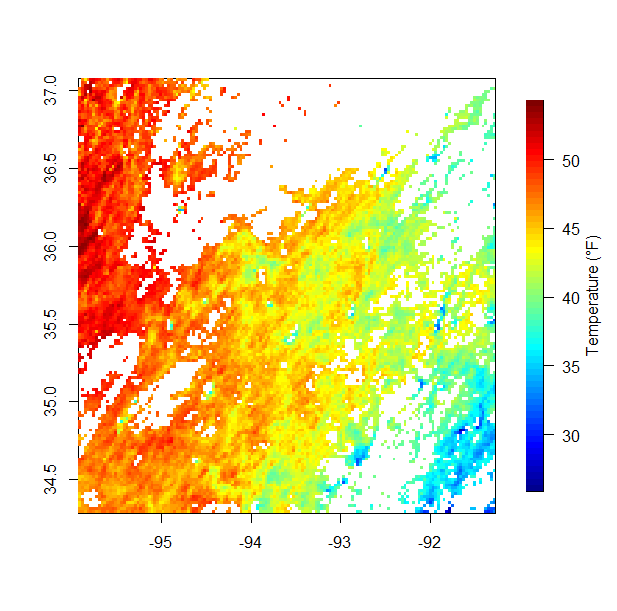} } 
    \caption{Terra temperature readings from \citep{Heaton2019}}
    \label{fig:heaton}
\end{figure}

Terra is an earth-observing satellite operated by NASA. \citep{Heaton2019} used daytime North American land surface temperature readings from Terra’s MODIS sensor to compare different spatial methods. The dataset has 148,309 observations and is partitioned into a training set of size 105,569 and a test set of size 42,740. The data is plotted in Figure \ref{fig:heaton}.

The University of California, Irvine (UCI) Machine Learning Repository contains datasets from a wide range of fields. The Elevators dataset describes an aeronautical experiment and consists of 16599 observations and 18 attributes. \citep{GPytorch} used this dataset to evaluate \texttt{GPytorch} for non-parametric regression. GP can be used for non-parametric regression by interpreting the attributes as locations and assuming constant mean.

\subsection{Model}
\label{sec:model}
We measure the performance of the software by fitting the following GP model on the three datasets:

\begin{equation}
Y= \beta + \eta +\epsilon \label{eq:model}
\end{equation} 

The response $Y=(Y_1,\ldots,Y_n)$ corresponds to wind speed for the Jason-3 data, temperature for Terra, and change in altitude for Elevators, with the location for $Y_i$ denoted as $Z_i=(Z_{i1}, \ldots,Z_{id})$. $\beta$ is the model intercept, $\eta\sim N(0,\Sigma(\sigma^2,\rho_1,\ldots,\rho_d))$ is the spatial stochasticity, and $\epsilon\sim N(0,\tau^2\sigma^2I_n)$ is the unstructured noise or nugget effect. The covariance matrix $\Sigma$ is defined by the exponential covariance function with variance $\sigma^2$ and range parameters $\rho_1\ldots\rho_d$.
\begin{equation}
Cov[Y_i,Y_j] = \sigma^2\exp\left(-\sqrt{\sum_{k=1}^d\left(\frac{Z_{ik}-Z_{jk}}{\rho_k}\right)^2}\right)
\end{equation}
When modeling the satellite data the range parameters are assumed to be equal $\rho_1=\cdots=\rho_d$, and the locations are embedded in $\mathbb{R}^3$ \citep{guinnessspheres}. For the Elevators data the range parameters are allowed to differ, and the attributes are used as the locations.

\subsection{Procedure}
\label{sec:procedure}

We compare the three implementations discussed in Section \ref{sec:methodology} by fitting Equation \ref{eq:model} on a subset of 50,000 points from the Jason-3 dataset. The dataset is divided into ten partitions and a cross-validation-like scheme is used to find the average runtime over the folds of size 45,000. The times for Fisher scoring are reported in Table \ref{tab:gpgpucomp} for $m=10,30,60$ neighbors and with a random permutation for the reordering. The block method could not be fit with $m=60$ neighbors because it exhausted all shared memory. The fastest method, thread-per-observation, is implemented in \texttt{GpGpU} and compared to \texttt{GpGp} and \texttt{GPytorch}.

We run \texttt{GpGp} on both a single-core and six-core system and use the highly optimized OpenBLAS linear algebra library. \texttt{GPytorch} allows for GPU acceleration but does not support Vecchia Approximation, so we instead use sparse Gaussian process regression (SGPR) and deep kernel learning with structured kernel interpolation (DKL+SKI). SGPR is an inducing point method that produces a low-rank representation of the covariance matrix. We use $w=300,600,1200$ inducing points, with the number of points limited by memory. DKL+SKI combines a neural network with inducing points placed on a grid to increase the flexibility of the model while inducing a covariance matrix with Toeplitz structure. We use the neural network architecture from \citep{GPytorch} and a grid of dimensions $100\times100$.

Due to memory constraints, the DKL+SKI model could not be fit on the Terra data and none of the \texttt{GPytorch} models could be fit on the full Jason-3 data. A cross-validation-like scheme was used to calculate the average runtime and RMSE on the Elevators, Jason-3 subset, and full Jason-3 data, while the Terra dataset was already divided into test and training sets so no cross-validation was performed. 

Runtime measures the time needed to fit the model, including data transfer from the CPU to the GPU. For \texttt{GpGpU} and \texttt{GpGp} the time needed to reorder the points and find the nearest neighbors is not included. Runtimes and RMSE for Elevators and Jason-3 are reported in Table \ref{tab:time} and Table \ref{tab:rmse} respectively, while the results for the Terra data are reported in Table \ref{tab:terra}.

\begin{figure}[h!]
\centering
\includegraphics[scale=0.6]{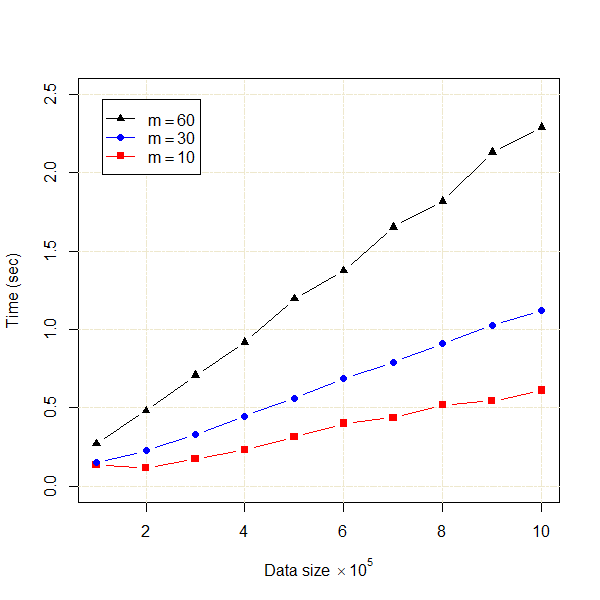}
\caption{Time (sec) for \texttt{GpGpU} with the thread-per-observation implementation to calculate the likelihood, gradient, and Fisher information of the GP model in Equation 5 for $m=10,30,60$ neighbors on subsets of the Jason-3 data with the exponential isotropic covariance function. This corresponds to one iteration of Fisher scoring, and includes all GPU overhead, such as memory allocation and memory transfers. Each time is an average of five runs.}\label{fig:timevsdata}
\end{figure}

\begin{table}[h!]
\caption{Time (sec) of fitting a GP model with three different GPU implementations of Vecchia Approximation on a subset of the Jason-3 data. The data is partitioned into $k=10$ folds and a cross-validation-like scheme is used to obtain the average runtime, with each fit performed on 90\% of the data, or $n=4.5\times 10^4$ observations. Fitting is Fisher scoring with the average number of iterations in parentheses. Time includes all overhead associated with the GPU, but not reordering the points or finding nearest neighbors. The models are fit with $m$ neighbors. The block method ran out of memory for $m=60$.}\label{tab:gpgpucomp}
\begin{tabular*}{\textwidth}{@{\extracolsep\fill}llll}
\toprule%
&$m=10$ & $m=30$ & $m=60$\\
\midrule
Thread &0.315 (8.1) &0.568 (7.5) &2.173 (9.2)\\
Block &0.522 (8.1)& 3.568 (7.5)& \cc \\
Batch &0.753 (8.1)& 2.846 (7.5)& 11.849 (9.2)\\
\botrule
\end{tabular*}
\end{table}

\begin{table}[h!]
\caption{Time (sec) of fitting a GP model on various datasets with \texttt{GpGpU} and \texttt{GPytorch}. Each dataset is partitioned into $k=10$ folds and a cross-validation-like scheme is used to obtain the average runtime, with each fit performed on 90\% of the data. \texttt{GpGpU} performs Fisher scoring for maximum 40 iterations and \texttt{GPytorch} uses the Adam optimizer for 100 iterations. The average number of iterations of Fisher scoring is in parentheses. The number of nearest neighbors is $m$ and the number of inducing points is $w$.  Reordering the points and finding the nearest neighbors is not included in the times.}\label{tab:time}
\begin{tabular*}{\textwidth}{@{\extracolsep\fill}lllll}
\toprule%
& \thead{Elevators \\($n=15$k)} &\thead{Jason-3 \\($n=45$k)} & \thead{Jason-3 \\($n=960$k)}&
          \thead{Jason-3 + Time \\($n=960$k)}\\
\midrule
GpGpU $(m=10)$          & 3.469 (40)& 0.315 (8.1) & 7.772 (7.9) & 10.279 (11)\\
        GpGp, 1 core $(m=10)$   & 22.527 (40)& 2.392 (8.1)       & 54.1485 (7.9)& 79.1318 (11)\\
        GpGp, 6 cores $(m=10)$  & 6.669 (40) & 0.728 (8.1)       & 12.565 (7.9) & 18.173 (11)\\
        GpGpU $(m=30)$          & 6.905 (37.6)      & 0.568 (7.5) & 53.123 (28.8) & 23.888 (12)\\
        GpGp, 1 core $(m=30)$   & 139.185 (37.6)& 10.427 (7.5)  & 174.166 (28.8) & 408.774 (12)\\
        GpGp, 6 cores $(m=30)$  & 26.725 (37.6)& 2.112 (7.5) & 894.821 (28.8) & 74.184(12)\\
        SGPR $(w=300)$          & 1.469     & 2.403 & \cc &\cc \\
        SGPR $(w=600)$          & 1.640    & 3.061  &\cc & \cc\\
        SGPR $(w=1200)$         & 2.293    & 4.750  & Out of memory \cc &\cc \\
        DKL+SKI                 & 29.517    & 78.069 & \cc  &\cc \\
\botrule
\end{tabular*}
\end{table}

\begin{table}[h!]
\caption{RMSE of a GP model fit on various datasets with \texttt{GpGpU} and \texttt{GPytorch}. Each dataset is partitioned into $k=10$ folds. Cross-validation is then used to obtain the average RMSE, with each fit performed on 90\% of the data and prediction on the remaining 10\%. \texttt{GpGpU} uses kriging with 60 neighbors while \texttt{GPytorch} uses the posterior predictive mean function. The number of nearest neighbors used for fitting the model is $m$ and the number of inducing points is $w$.}\label{tab:rmse}
\begin{tabular*}{\textwidth}{@{\extracolsep\fill}lllll}
\toprule%
& \thead{Elevators \\($n=15$k)} &\thead{Jason-3 \\($n=45$k)} & \thead{Jason-3 \\($n=960$k)}&
          \thead{Jason-3 + Time \\($n=960$k)}\\
\midrule
GpGpU $(m=10)$ & 5.3            & 2.230 & 3.181 &1.281 \\
        GpGp, 1 core $(m=10)$& 5.3      & 2.230 & 3.181 & 1.281\\
        GpGp, 6 cores $(m=10)$ & 5.3    & 2.230 & 3.181 & 1.281 \\
        GpGpU $(m=30)$ & 5.295          & 2.227 & 3.181 & 1.281\\
        GpGp, 1 core $(m=30)$ & 5.295   & 2.227 & 3.181 & 1.281\\
        GpGp, 6 cores $(m=30)$ & 5.295  & 2.227 & 3.181 & 1.281\\
        SGPR $(w=300)$ & 6.710          & 2.961 & \cc &\cc \\
        SGPR $(w=600)$ & 6.709          & 2.914 &\cc & \cc\\
        SGPR $(w=1200)$ &6.700          & 2.874   & Out of memory \cc &\cc \\
        DKL+SKI & 6.665                 & 3.449  & \cc &\cc \\
\botrule
\end{tabular*}
\end{table}

\begin{table}[h!]
\caption{Time (sec) and RMSE of fitting a GP model on the Terra data with \texttt{GpGpU} and \texttt{GPytorch}. The data is partitioned into a training and test set as in \citep{Heaton2019}.}\label{tab:terra}
\begin{tabular*}{\textwidth}{@{\extracolsep\fill}llll}
\toprule%
&Time (sec) & RMSE & Iterations\\
\midrule
GpGpU$(m=10)$ & 2.289&1.370 &18\\
        GpGp-1$(m=10)$ & 11.967&1.370&18\\
        GpGp-6 $(m=10)$ & 2.673 &1.370&18\\
        GpGpU $(m=30)$ & 2.842& 1.468&13\\
        GpGp-1 $(m=30)$ &41.452 & 1.468&13\\
        GpGp-6 $(m=30)$ &7.528& 1.468&13\\
        SGPR $(w=300)$ &4.879 &3.022 \\
\botrule
\end{tabular*}
\end{table}

\subsection{Results} \label{sec:results}

The thread-per-observation method is faster than the block-per-observation and batched methods, which we attribute to minimal synchronization and the low latency of registers. Further examination shows that it also scales well with $m$ and $n$, as seen in Figure \ref{fig:timevsdata}.

We find that \texttt{GpGpU}, using the thread-per-observation method, is always faster than \texttt{GpGp}, with fitting speedups as high as 17 times over single core and 3 times over six-core. \texttt{GpGpU} is faster than \texttt{GPytorch} on the geospatial data, fitting the Jason-3 subset nearly 200 times faster than DKL+SKI and 6 times faster than SGPR while having a lower RMSE. It also uses less memory, being able to fit the full Jason-3 data while SGPR and DKL+SKI cannot. \texttt{GpGpU} is slower than \texttt{GPytorch} on the Elevators data, but achieves better predicitive accuracy. \citep{GPytorch} fit the Elevators data with a Mat\'ern with $\nu=\frac{5}{2}$, but we found higher predictive accuracy with the exponential isotropic covariance. 

\begin{figure}[h!]
\centering
\includegraphics[scale=0.6]{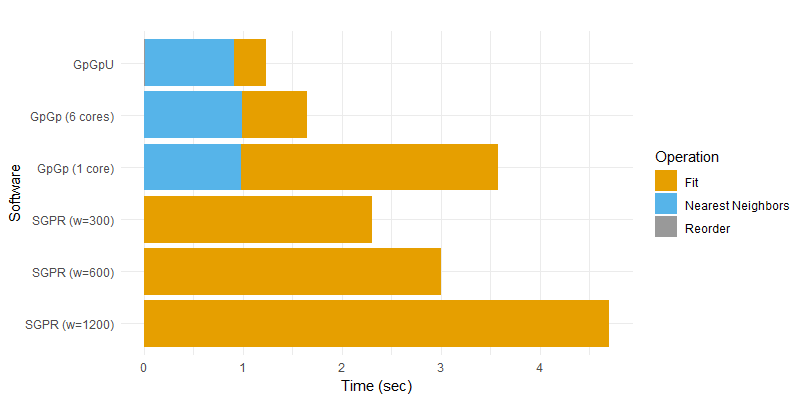}
\caption{Time profile of fitting GP on $4.5\times10^5$ Jason-3 observations with different software. The Vecchia Approximation software use $m=10$ neighbors. The SGPR methods use $w$ inducing points.}\label{fig:profile}
\end{figure}

\texttt{GpGpU} remains competitive even when the time for reordering the points and finding nearest neighbors is considered, in that it is still faster than \texttt{GPytorch} on the geospatial dataset. The nearest neighbors subroutine runs sequentially on the CPU and takes longer than the Fisher scoring, while the reordering of the points takes minimal time. A full time profile of fitting a GP on the Jason-3 subset is in Figure \ref{fig:profile}.

\section{Discussion} \label{sec:discussion}

\texttt{GpGpU} addresses the need for software that can quickly fit large geospatial GP models and provides a comparison of three different approaches. Our new algorithm leverages the insight that Vecchia Approximation fully parallelizes each iteration of Fisher scoring while maintaining a low memory burden for each parallel process. This allows intermediary values to be stored in registers, which have the lowest latency. As a result, our method differs from most other GP software that rely on global memory and do not explore how memory type affects performance.

Our \texttt{GpGpU} package outperforms \texttt{GpGp} and \texttt{GPytorch} on large geospatial datasets and is competitive on higher-dimensional data. We did not compare \texttt{GpGpU} to \citep{pan2024} because that paper only reported times for calculating the likelihood, while we optimized our code to fit the model as quickly as possible, which includes calculating the likelihood, gradient, and Fisher information together. \citep{pan2024} did not provide code for calculating the gradient and Fisher information.

Our implementation has several limitations, including possibly exhausting memory when the model has a large number of linear predictors or covariance parameters. Due to CUDA's lack of support for the modified Bessel function of the second kind, we cannot use the popular Mat\'ern covariance function with arbitrary smoothness. However, we believe our software is still of great use to practitioners, as it is able to support large geospatial datasets and many other covariance functions.

Possible future improvements to \texttt{GpGpU} include parallelization over multiple GPUs, supporting arbitrary covariance functions with automatic differentiation, and grouping observations. We also hope to create GPU algorithms for important subroutines, specifically finding the nearest neighbors, which was shown to take longer than optimizing the loss function.

\section*{Software}

The \texttt{GpGpU} package is available at \url{https://github.com/zjames12/GpGpU}.



\begin{appendices}


\end{appendices}

\bibliography{sn-bibliography}

\end{document}